\documentclass[journal]{IEEEtran}

\usepackage{cite}
\usepackage{amsmath}
\DeclareMathOperator*{\argmin}{arg \, min}
\renewcommand{\vec}[1]{\boldsymbol{\mathbf{#1}}} 

\usepackage{xcolor} 

\usepackage{algorithm}
\usepackage{algpseudocode}

\usepackage{standalone} 

\usepackage{diagbox} 
\usepackage{multirow} 
\usepackage{makecell} 

\usepackage{array}
\usepackage[caption=false,font=footnotesize]{subfig}

\begin{document}
\title{Multiplant Nonlinear System Identification by Block-Structured Multikernel Neural Networks in Applications of Interference Cancellation}
%
%

\author{Svantje~Voit,~\IEEEmembership{Student Member,~IEEE, }%
        and~Gerald Enzner,~\IEEEmembership{Senior~Member,~IEEE}
\thanks{This work was supported by grant DFG EN 869/4-1, project no. 449601577.}
\thanks{}
}

\markboth{}%
{}
%


\maketitle

\begin{abstract}
  Problems of linear system identification have closed-form solutions, e.g., using least-squares or maximum-likelihood methods on input-output data. However, already the seemingly simplest problems of nonlinear system identification present more difficulties related to the optimisation of the furrowed error surface. Those cases include the Hammerstein plant with typically a bilinear model representation based on polynomial or Fourier expansion of its nonlinear element. Wiener plants induce actual nonlinearity in the parameters, which further complicates the optimisation. Neural network models and related optimisers are, however, well-prepared to represent and solve nonlinear problems. Unfortunately, the available data for nonlinear system identification might be too diverse to support accurate and consistent model representation. This diversity may refer to different impulse responses and nonlinear functions that arise in different measurements of (different) plants. We therefore propose multikernel neural network models to represent nonlinear plants with a subset of trainable weights shared between different measurements and another subset of plant-specific (i.e., multikernel) weights to adhere to the characteristics of specific measurements. We demonstrate that in this way we can fit neural network models to the diverse data which cannot be done with some standard methods of nonlinear system identification. For model testing, the subset of shared weights of the entire trained model is reused to support the identification and representation of unseen plant measurements, while the plant-specific model weights are readjusted to specifically meet the test data.
\end{abstract}


\section{Introduction}
%
%
%

\IEEEPARstart{I}{nterference} cancellation typically relies on plant identification in order to duplicate and compensate undesired interference from a primary observation signal that also contains an information-bearing desired signal. The goal is to enhance the accessibility of the information in the desired signal by subtracting an estimated interference from the primary signal. To do so, the primary signal serves as a target response for an interference plant model. An auxiliary signal, which is ideally independent from the information-bearing signal, serves as the model input. The model parameters (in some cases termed filter coefficients) are then adjusted in a system identification fashion to match the model output to the actual interference linked to the same auxiliary signal by the initially unknown plant. The model structure must be architected for its capability of representing the plant, while the actual plant structure is assumed to be available as domain-knowledge.

The process of interference cancellation is crucial in applications like acoustic echo cancellation (AEC) in hands-free systems for speech communication \cite{Benesty2001, Haensler2006,VaryMartin2006, enzner2014acoustic} or self-interference cancellation (SIC) in full-duplex radios for wireless communication \cite{Heino_2015,Herd_2019,Smida2023, sexton20175g,He2022frequency}. In hands-free speech communication, a microphone not only receives the desired speech from the near end, but also an acoustic echo of the far-end voice. The cancellation of this echo signal by estimating the echo path model is required as an important subtask in acoustic echo control systems. Similarly, inband full-duplex radios simultaneously transmit and receive through the same time and frequency resource. The desired signal from a remote side thus cannot be reliably restored at the receiver since the signal emitted by the transmitter strongly penetrates the receiver chain as self-interference. Very accurate cancellation of the self-interference is required here.

In both cases, the nonlinear (NL) plant, i.e., the echo or self-interference path, impedes the modelling of the interference process that needs to be compensated. 
In the acoustic system, high playback volumes on small transducer geometries cause the loudspeaker to introduce nonlinear behaviour \cite{klippel_nonlinear_1992}. In wireless systems, nonlinearities can be found in terms of power amplifiers \cite{morgan2006generalized} or analogue-to-digital converters \cite{ahmed2015all} with saturation or clipping behaviours. As these applications exhibit nonlinear aspects that need to be modelled as accurately as possible to achieve the desired cancellation, the use of classical linear system identification models would be limited to insufficient results \cite{Mossi_nonlinear_2010, Enzner_SICdata_Arxiv_2024}.

The task of nonlinear system identification has been approached by different model classes. Nonlinear stochastic state-space models are a very general framework applied to the AEC problem, where the nonlinearity is captured by sampled distributions \cite{Huemmer2018} or functional expansions \cite{Malik2012,Malik2013} around an acoustic system state. Another very general class is given by nonlinear autoregressive moving average models with exogenous inputs (NARMAX) \cite{billings_nonlinear_2013,chen1990practical}. Here, the model is a general function on a sequence of past inputs, outputs and noise terms. It is initially formulated in a very abstract way before being further specified as a more concrete model the parameters of which can be estimated. If domain-knowledge allows, the model architecture can be reduced to a subsystem of the general NARMAX framework, whereby, e.g., the Volterra model \cite{zeller2009fast}, the block-structured models, such as Wiener \cite{wigren1994convergence} or Hammerstein \cite{greblicki1989nonparametric}, memory polynomials \cite{morgan2006generalized, makela2003effects}, and many neural network architectures \cite{chen1990non, kelley2024comparison} can be regarded as special cases.

When dealing with applications, such as AEC and SIC, domain-knowledge is always provided to justify more specific model architectures, paving the way for a physically motivated design to localise both linear and nonlinear system components and allowing for systematic evaluation of the components. However, there is always a trade-off between algorithmic complexity and modelling capability \cite{Malik2012,Malik2013, Huemmer2018,Haykin2002,Stenger1999,Stenger2000,guerin2003, kuech2006, malik2011fourier,ding2004robust}. Volterra series, for instance, can represent a variety of nonlinear behaviours, but are associated with large computational effort hampering practical relevance. An implementation of Volterra series higher than second order may be unrealistic, although it might be physically necessary. Models are therefore often restricted to a memory element before or after a memoryless nonlinearity, i.e., models consisting of a static nonlinear block surrounded by one or two dynamic linear blocks. 
These block structures are called Wiener, Hammerstein or Wiener-Hammerstein, depending on the respective block order.

Neural networks can be a powerful tool for redesigning the trade-off, as the simple Wiener or Hammerstein nonlinearity can be easily represented and extended by creating more complex models bottom-up when further modelling capability is required according to more specific domain-knowledge. 
Neural networks were used in applications where system identification is desirable, but the networks were not configured so as to meet the system identification purpose. In some cases, not only the auxiliary reference signal is fed into the model, which is necessary to enforce nonlinear plant representation, but also the primary observation signal appears as an input \cite{Ivry2021, braun2022task, seidel20212, Zhang2022, Westhausen2021}. These ''look-alikes'' of plant modelling do not enforce a decorrelation of the network output from the desired signal, since the desired signal is part of the network input via the primary signal. Subtraction of the model output signal from the primary signal can therefore distort the desired signal during the interference cancellation.

The public availability of realistic datasets further supports the development of neural networks for nonlinear modelling. For realistic experiments in AEC applications, data is provided within the popular AEC challenge database \cite{Sridhar2021}. For the wireless application, the WLAN toolbox \cite{Gast2012_802_11n} allows the simulation of realistic data \cite{Enzner_SICdata_Arxiv_2024} for the development of SIC models. In both domains, the underlying plants can be considered as a block-structured transmission path consisting of different linear and nonlinear components.
The data shows that plant diversity remains a problem in the use of neural networks and, to our knowledge, has not yet been fully addressed. The desired variability w.r.t.\ different plants is achieved in \cite{Halimeh2019} by replacing parts of the network with an adaptive filter in the test mode. For network training, however, the data must be constrained to consistent room impulse responses therein.
The system in \cite{birkett1995acoustic}, supposedly, determines a variable network model individually for each given input signal, i.e., instead of identifying a general nonlinearity across all plants, a fully plant-specific identification is performed. 

In this paper, we introduce a framework to represent shared and variable plant behaviour in the training and evaluation of system identification neural networks. By leveraging domain knowledge of the plant, we propose multikernel neural network models with a subset of its trainable weights shared across the data, while another set of multikernel weights is adjusted for plant-specific representation. We initially validate the framework with white-noise data and further apply it to more realistic problems, such as AEC with Hammerstein nonlinearity and SIC with Wiener nonlinearity.

The paper is organised as follows: Section \ref{sec:ProblemFormulation} describes the general cancellation problem. Section \ref{sec:BSNLS} outlines the plant-specific diversity of block-structured nonlinear systems. Section \ref{sec:NFA} reviews classical methods for nonlinear modelling as well as the neural-network approximation. Plant-specific modelling via the multikernel approach is introduced in Section \ref{sec:FIRblock}, which is then combined in Section \ref{sec:model} with shared nonlinear modelling to form multikernel neural networks, the functionality of which is also demonstrated in this section. Sections \ref{sec:AEC} and \ref{sec:SIC} examine the approach in specific areas of application, i.e., the  acoustic echo cancellation and the self-interference cancellation, respectively. Section \ref{sec:conclusion} concludes.

\section{Cancellation Problem Formulation}
\label{sec:ProblemFormulation}
\begin{figure}[!t]
  \centering
  \includegraphics[width=1\columnwidth]{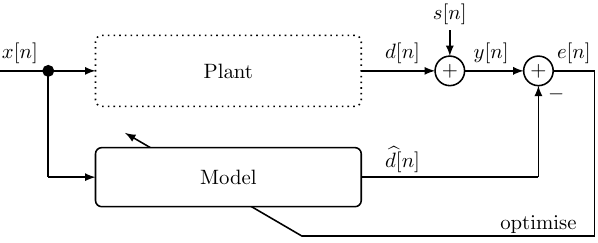}
  \caption{Interference cancellation using a model based on system identification.}
  \label{fig:SysId}
\end{figure}
In interference cancellation, a model is employed to identify and cancel an undesired interference $d[n]$ present in a primary signal $y[n]$, which also contains an information-bearing signal component $s[n]$. 
Fig.~\ref{fig:SysId} depicts such a setup.
Here, the primary signal $y[n] = d[n] + s[n]$ is used as the desired response of the model. An auxiliary signal $x[n]$ serves as the model input, whilst this auxiliary signal is obtained from a reference sensor which is designed such that the information-bearing signal in the primary signal is not detectable via the reference signal and is therefore physically independent.

The method of interference cancellation structures and adjusts weights of the model so that its output $\widehat{d}[n]$ at discrete time $n$ matches the interference $d[n]$, which is linked to the auxiliary reference signal $x[n]$ through an unknown plant. 
This method can be understood as system identification in which the plant is identified using a most appropriate model. Here, potential nonlinearities in the system must be included in the model structure, such that the interference cancellation is able to subtract the estimated interference $\widehat{d}[n]$ from the primary signal $y[n]$ with best fidelity to $\widehat{d}[n]$. 
The optimal result is an error signal $e[n] = y[n] - \widehat{d}[n]$ that conveys only the desired signal $s[n]$, effectively removing any component related to $x[n]$.
This process of interference cancellation is crucial in applications where isolating the interference signal is vital for further processing or analysis.

\section{Block-Structured Nonlinear Multiplants}
\label{sec:BSNLS}
Domain-knowledge is the basis for designing model-based prototypes of plant structures that consist of dynamic linear and static nonlinear blocks, which may appear in different constellations for certain application areas. A dynamic block represents linear-time-invariant (LTI) behaviour, i.e., typically with memory, which can be characterised by an impulse response model of the block. Considering different realisations or measurements of plants to be formally represented in this paper as multiplants, we introduce a plant-specific impulse response ${h}^{[\kappa]}[n]$ for a dynamical block of plant $\kappa$. Here, the plant index $\kappa$ depicts an additional dimension of system representation beyond classical LTI theory. Our intention hereby is to describe significant system variability across different observations (i.e., sample data in the neural network context) while assuming LTI behaviour per observation. Static blocks then refer to memoryless nonlinear functions $f^{[\kappa]}(x[n])$ for which it is not possible to describe a convolutive input-output relationship or commutativity with its input $x[n]$. As before, the dimension $\kappa$ shall here represent potential variability of the respective block across sample data of a plant structure, while the nonlinear function is still assumed to be fixed within a single observation sequence of plant $\kappa$.

For both block-types, we will distinguish between variant and invariant blocks. While those variant blocks or variant plants made from such blocks were indexed with the $\kappa$, their invariant counterparts will be denoted as dynamical blocks with impulse responses ${h}^{[\kappa]}[n] = {h}[n]$ or static blocks with nonlinear function $f^{[\kappa]}(x[n])= f(x[n])$ across all samples $\kappa = 1, \dots , K$ of a multiplant. 
Consequently, invariant blocks can be understood as a special case of variant ones.

\begin{figure}[!t]
  \centering
  \subfloat[Wiener structure.]{\includegraphics[width=0.75\columnwidth]{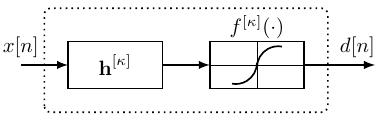}%
  \label{fig:WienerSysId}}
  \hfil
  \subfloat[Hammerstein structure.]{\includegraphics[width=0.75\columnwidth]{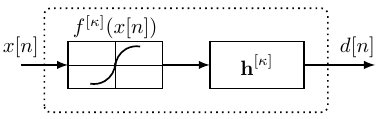}%
  \label{fig:hammersteinSysId}}
  \caption{Plants with different nonlinear block-structures.}
  \label{fig:sysId}
\end{figure}

In order to contrast with block structures, we recall the Volterra series expansions as a very abstract and contained structure to model a plant. It can form a variety of nonlinear functions, but potentially requires a huge number of parameters, which is increasing superlinearly with increasing memory and polynomial order, and is therefore limited to applications with low system orders \cite{Schetzen2006,mathews2000}. The following provides an overview of common block-structured nonlinear models with a deliberately small number of parameters and restricted application based on dedicated domain knowledge.

The \textit{Wiener structure} \cite{Schetzen2006,mathews2000} is defined as a dynamic block followed by a static nonlinear block to create an overall model of nonlinearity with memory. In addition, a distinction is made between cascaded and parallel \cite{birkett1995acoustic,morgan2006generalized} Wiener structures, whereby the cascaded structure is a special case of the parallel structure. The parallel structure results as a sum of multiple Wiener structures \cite{ding2004robust}, therefore, if the entire structure is composed of only one Wiener subsystem, it results in the cascaded version. It is shown in Fig.~\ref{fig:WienerSysId}, where the input $x[n]$ is first processed by convolution related to the dynamic block with impulse response $h^{[\kappa]}[n]$ and then by the static nonlinear function $f^{[\kappa]}(\cdot)$ such that the output signal reads
\begin{equation} \label{eq:wiener}
  d[n] = f^{[\kappa]}\!\left(\sum_{m=0}^{L}h^{[\kappa]}[m] \,x[n-m]\right) \; .
\end{equation}
Wiener modelling is a seemingly simple method of combining memory and nonlinearity, however, its nonlinearity in the parameters challenges the system identification \cite{wigren1999compensation, nordsjo2001identification}.

The \textit{Hammerstein structure} consists of a static nonlinearity followed by a dynamical block \cite{nelles2020,narendra1966,nordsjo2001} and thus can be linear in the parameters (if the nonlinear function is predetermined). Parallel \cite{Malik2012} and cascaded \cite{Malik2013} Hammerstein structures exist in the applications, whereby parallel structures are formed as a summation of multiple Hammerstein subsystems, enabling the representation of nonlinearity with memory. A cascaded Hammerstein structure employs the same linear dynamical block for each subsystem and thus collapses to the modelling of nonlinearities without memory. It is shown in Fig.~\ref{fig:hammersteinSysId} and, according to the commutation of convolution and nonlinear mapping compared to the Wiener structure, the output signal of Hammerstein structures differently reads 
\begin{equation} \label{eq:hammerstein}
  d[n] = \sum_{m=0}^{L}h^{[\kappa]}[m] \,f^{[\kappa]}\left(x[n-m]\right) \; .
\end{equation}
With its linearity in the parameters, it can more easily be employed with linear estimation algorithms \cite{Haykin2002, kim2001digital}.

There are as well mixed forms of Wiener and Hammerstein structures, i.e., \textit{Wiener-Hammerstein structures}, that for example consist of a static block sandwiched between two dynamic blocks. This paper, however, will not cover all  possibilities of domain-specific plant representation.

\section{Nonlinear Function Representation}
\label{sec:NFA}
In none of the structures, the actual nonlinear function is typically known a-priori in closed form. Therefore, a parametric representation ${f}(x[n]; a_p,  b_p)$ is needed to set up and fit the function. Before, the terminology of functions being linear or nonlinear in the parameters is briefly clarified. To this end, a functional form
\begin{equation} \label{eq:nlapprox}
  {f}(x[n]; a_p,  b_p) = \sum_{p=1}^{P} a_p \Phi_p ( x[n] ; \, b_p) \; 
\end{equation}
is considered, where the function output is determined via the input $x[n]$, the parameters $a_p$ and $b_p$, and the nonlinear basis functions $\Phi_p(\cdot)$. The function ${f}(x[n]; a_p,  b_p)$ is then linear in its parameters $a_p$ that directly relate to the function output, while nonlinear in the parameters $b_p$ of the basis functions.

\subsection{Representations with Linearity in the Parameters}
\label{subsec:LP}
Parametric expansions can be specified by fixed basis functions $\Phi_p(\cdot)$ weighted by linear coefficients $a_p$, while any nonlinear coefficients $b_p$ are discarded.

One important representation is the power series with primitive polynomials of $x[n]$ as the basis functions,
\begin{equation} \label{eq:powerApprox}
    {f}_{\text{power}}(x[n]; \, a_p) = \sum_{p=1}^{P} a_p x^p [n]\;,
\end{equation}
typically assuming the range $-1<x[n]<1$ for stability, and the parameters to be estimated are the $p$-th coefficients $a_p$ of the $P$-th order polynomial. Optimality of the coefficients may be designated in terms of the least-squares error between the parametric representation and the actual nonlinear function:
\begin{equation}
    \widehat{a}_p = \argmin_{a_p} \int_{-1}^{+1} \left( {f}(x[n]) - \sum_{p=1}^{P} a_p x^p [n] \right)^{\!\!2} \, d x[n] \; .
\end{equation}

Another representation relies, for instance, on the odd Fourier series with orthogonal sinusoidal basis functions,
\begin{equation}
    {f}_{\text{fourier}}(x[n]; \, a_p) = \sum_{p=1}^{P} a_p \sin \left( 2\pi \cdot p \cdot \frac{x[n]}{T_x} \right) \; ,
\end{equation}
with a hyperparameter $T_x$ defining the fundamental period (i.e., effectively the $x[n]$ range) of a periodical function. The linear parameters for least-squares functional approximation are known as the odd Fourier coefficients
\begin{equation}
    \widehat{a}_p = \frac{2}{T_x} \int_{-T_x/2}^{T_x/2}  {f}(x[n]) \cdot \sin\! \left( 2\pi \cdot p \cdot \frac{x[n]}{T_x} \right) \, d  x[n] \; .
\end{equation}

Whether the basis functions qualify for good representation depends on the target function $f(x[n])$, on the input $x[n]$, specifically the adequate range and the distribution of the input, and on the manageable nonlinear order $P$.

\subsection{Representation with Nonlinearity in the Parameters}
\label{subsec:NLP}

In our applications we would require arbitrary placement of a nonlinear function in a block structure and we would have to cope with potentially uncertain input signal range. We thus like to have control over the range of the input signal before the nonlinearity takes place and therefore introduce a nonlinear function representation by neural networks, specifically the multi-layer perceptron (MLP) \cite{bishop2006pattern, goodfellow2016deep}. It consists of $\ell=1,...,D-1$ hidden layers each with $p_\ell=1,...,P_\ell$ output channels and nonlinear input-output relation in the parameters,
\begin{equation} \label{eq:hidden_layer}
     {f}_{p_{\ell+1}}[n] =  \tanh\! \left( \sum_{p_\ell=1}^{P_\ell} b_{p_{\ell+1},p_\ell}\; {f}_{p_\ell}[n] \right) \; ,
\end{equation}
with the first layer input ${f}_{p_0}[n]=x_{p_0}[n]$, $p_0=1,...,P_0$, and $P_0=I$ denotes available input channels. The typical $\tanh$-activations conclude the arithmetic logic of each hidden layer. For regression 
a linear output layer may finally aggregate the available nonlinear representations of the last hidden layer as
\begin{equation} \label{eq:output_layer}
    {f}_p\!\left( x[n]; \, a_{p,p_D}, b_{p_{\ell+1},p_\ell}\right) =  \sum_{{p_D}=1}^{P_D} a_{p,p_D}\, {f}_{p_D}[n] \; . 
\end{equation}

\begin{figure}[!t]
  \centering
  \includegraphics[width=0.5\columnwidth]{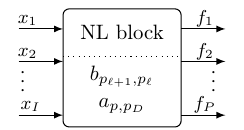}
  \caption{Nonlinear memoryless block which maps an $I$-dimensional input to a nonlinear $P$-dimensional output using internal $\tanh$-activations.}
  \label{fig:NLblock}
\end{figure}

The MLP-based nonlinear block is thus denoted as these stacked $D+1$ layers with its trainable parameters $ a$ and $ b$, the input dimensionality $I$, and output dimensionality $P$ as shown in Fig.~\ref{fig:NLblock}. 
This nonlinear (NL) block can be well placed in arbitrary positions of a model structure as the nonlinear parameters $ b$ are capable of controlling the block's input and the linear parameters $ a$ its output, respectively.

Our implementation of this NL block (using PyTorch functions) employs $D+1$ layers of 1D-Conv specification with kernelsize of one for memoryless nonlinearity. In this case, the padding does not matter. The input-tensor $\mathcal{X}$ then consists of three dimensions $(K , I , M)$, where $K$ denotes different realisations of a plant structure, $I$ the input-channel dimensionality, and $M$ the number of time samples of an input sequence $x[n]$. For a single-channel time-domain input signal it holds $I=1$. According to \eqref{eq:hidden_layer} and \eqref{eq:output_layer}, the NL block then maps the input to an output tensor ${\mathcal{F}}$ with dimension $(K , P,  M)$. Here, the second tensor dimension forms a $P$-th order nonlinear expansion. The first tensor dimension remains untouched for parallel processing and overall loss minimisation. The block's trainable weights $ a$ and $ b$ are thus shared for system identification across the different plant observation.

\section{Linear Multikernel Representation}
\label{sec:FIRblock}
We briefly recap that Section~\ref{sec:BSNLS} has introduced the dimension of plant variability in problems of nonlinear system identification. Section~\ref{sec:NFA} has described the possibility of neural network modelling to represent the nonlinear dimension in those problems of nonlinear system identification. Given the trend of large datasets for the optimisation of neural networks, the purpose of this upcoming section is to address the issue of plant variability within a dataset by describing a corresponding neural network representation. Specifically, the idea is to oppose to the dimension of plant variability the dimension of multikernel representation in the networks. The multikernel approach thus invokes multiple specific kernels to represent the specific plants of specific data samples. However, this multikernel representation is not necessarily devoted to an entire model architecture under consideration for the system identification problem. Here in this section we thus introduce the multikernel approach only for the linear (yet important for system modelling) layers of an entire model. 

\subsection{Time-Domain FIR Block}

\begin{figure}[!t]
  \centering
  \includegraphics[width=0.5\columnwidth]{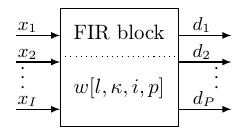}
  \caption{Multikernel FIR-Block which maps a $I$-dimensional input on a $P$-dimensional output to model LTI behaviour with memory $L$.}
  \label{fig:FIRblock}
\end{figure}

LTI behaviour of a plant component with input $x[n]$ can be modelled by FIR filtering with impulse response $w[l]$ at lag indices $l=0,...,L-1$. The corresponding input-output relationship of the plain FIR model at discrete time step $n$ is then written by convolution as simple as
\begin{equation}
        {d}[n] = \sum_{l=0}^{L-1}  x[n-l] \, w[l] \; ,
\end{equation}
with one-dimensional signal $x[n]$ and one-dimensional convolutional kernel $w[l]$. For versatile representation in the neural network context, several additional dimensions are required. In order to support system identification with $K$ different plants or different observations of a plant, a two-dimensional input sequence $x[n,\kappa]$ and a two-dimensional kernel $w[l,\kappa]$ of shape $(L,K)$ are required for representation. The input-output relationship is
then given by
\begin{equation}
        {d}[n,\kappa] = \sum_{l=0}^{L-1}  x[n-l,\kappa] \, w[l,\kappa] \;, 
\end{equation}
where the plant index $\kappa$ reproduces from the input to the output. For multiple-input multiple-output (MIMO) representation of different plants, we require a three-dimensional input signal $x[n,\kappa,i]$ and a four-dimensional kernel $w[l, \kappa, i, p]$ of shape $(L,K,I,P)$ to deliver a three-dimensional output as
\begin{equation}
        {d}[n, \kappa, p] = \sum_{i=1}^{I}\sum_{l=0}^{L-1}  x[n-l, \kappa, i] \, w[l, \kappa, i, p] \; ,
\end{equation}
where $I$ and $P$ here denote the multiple-input and multiple-output dimensionality, respectively. Further considering a segmentation of the input sequences according to  
\begin{equation}
        x[t, m, \kappa ,i] = x[tR + m, \kappa, i]\; ,
        \label{eq:frames}
\end{equation}
with $m = 0, \dots, M-1$, the new time step index of the subsequences or ''frames'', $M$ the frame length, $t$ the frame index, $R$ the frame shift, and thus $M-R$ the frame overlap, we got a four-dimensional input signal $x[t, m, \kappa ,i]$. It still uses the four-dimensional kernel to produce the output signal   
\begin{equation}
\label{eq:multikernelConv}
        {d}[t, m, \kappa, p] = \sum_{i=1}^{I}\sum_{l=0}^{L-1}  x[t, m-l, \kappa, i] \, w[l, \kappa, i, p] \; , 
\end{equation}
where the frames $t$ are reproduced from input to output with shared kernels according to the idea of the same plant being responsible for one input sequence and the frames therein. In order to obtain $m=0,...,R$ valid and seamless output time steps of the convolution for the given frame-shift $R$, the kernelsize must be constrained to $L=M-R+1$. 

As a baseline for the proposed multikernel model, for the sake of clarity, we write out a single-kernel model, i.e.,
\begin{equation}
        {d}[t, m, \kappa, p] = \sum_{i=1}^{I}\sum_{l=0}^{L-1}  x[t, m-l, \kappa, i] \, w[l, i, p] \; , 
\end{equation}
where the plant index is dropped from the multikernel in (\ref{eq:multikernelConv}), while it persists with input and output sequences. This essentially means that the kernel $w[l, i, p]$ is conventionally shared across the data samples corresponding to different plants, which, supposedly, hampers minimum mean-square error representation of data samples by the model.

\subsection{Frequency-Domain FIR-Block Representation}

Based on the success story of frequency-domain representations for adaptive online learning of FIR filters \cite{Haykin2002}, specifically in the field of speech processing \cite{Benesty2001,Enzner06}, this method is here adopted with the hypothesis of potentially advanced learning in the context of neural-network representation. We therefore provide a kernel definition with zero padding in the convolutive dimension of the original temporal time steps,
\begin{equation}
    w_z[m, \kappa, i, p] = \left\{ 
    \begin{array}{ll}
         w[m, \kappa, i, p] & \quad m=0,...L-1 \\
         0 & \quad m=L,...M-1\;,
    \end{array}
    \right.
\end{equation}
of kernel shape $(L+R-1,K,I,P)$ with the previous constraining to $M=L+R-1$. Its $M$-dimensional representation in the discrete Fourier transform (DFT) domain is
\begin{eqnarray}
\label{OLS_constrain}
    W[k, \kappa, i, p] & \!\!=\!\! & \sum_{m=0}^{M-1} w_z[m, \kappa, i, p] \, \mathrm{e}^{-\jmath2\pi m k / M} \\
    & \!\!=\!\! & \sum_{l=0}^{L-1} w[l, \kappa, i, p] \, \mathrm{e}^{-\jmath 2\pi l k / M}
\end{eqnarray}
with discrete frequency index $k=0,...,M-1$. Further using the input signal in the DFT domain along the $m$-dimension,  
\begin{equation}
\label{eq:fftInput}
    X[t,k,\kappa,i] =  \sum_{m=0}^{M-1} x[t,m,\kappa,i] \mathrm{e}^{-\jmath 2\pi m k / M} \; ,
\end{equation}
where merely the time step index $m$ is converted into the discrete frequencies $k$, we can obtain the predicted output signal in the time domain by elementwise spectral multiplication of matching dimensions in the DFT domain and inverse DFT,
\begin{eqnarray}
\label{weights_complex}
\label{eq:spectralMult}
        {D}[t,k,\kappa,p] & = & \sum_{i=1}^{I}  X[t,k,\kappa,i] \, W[k, \kappa, i, p]  \\
\label{eq:ifftOLS}        
        {d}[t,m,\kappa,p] & = &\frac{1}{M} \sum_{k=0}^{M-1} {D}[t,k,\kappa,p]\, \mathrm{e}^{+\jmath2\pi mk / M} \; , 
\end{eqnarray}
with valid output samples only for $ m=M-R,\dots,M-1$ according to the principles of overlap-save processing.

\subsection{FIR Block Implementation}

In our PyTorch implementation, the data passed through an FIR block is arranged as a four-dimensional input tensor. The first dimension is the batchsize and it hosts the frames $t=0,...,T$ for parallel processing with same kernels for one and the same plant index $\kappa$. The second dimension keeps the individual plants $\kappa=0,...,K$. Then follows the dimension of inputs $i=1,...,I$, before the last dimension is used to represent the sequence of time steps $m=0,...,M$ per frame. The FIR block thus converts a $(T, K, I, M)$-dimensional input-tensor $\mathcal{X}$ to a $(T, K, P, R)$-dimensional output tensor ${\mathcal{D}}$. The implementation of the FIR block, however, depends on the previous time- or frequency-domain logic.

\subsubsection{Time-Domain FIR-Block}

To form the system output with $P$ output channels, a number of $P$ 2D-Conv-Layers is set up in parallel. This extra effort is rooted in the fact that the output-channel dimension is employed for group processing in order to enable the representation of individual kernels for the $K$ individual plants (PyTorch: 'groups $=K$') which equals the number of input and output channels. We thus have specific kernels of size $(I , L)$ for each plant in the data. The padding regarding the dimension of time steps is set to 'valid'. Accordingly, the kernel slides along the dimension of time steps $m$ while aggregating the input dimension $i$.

\subsubsection{Frequency-Domain FIR-Block}

Complex-valued weights $W$ according to (\ref{weights_complex}) are set up with two real-valued weight tensors of size $(M,K,I,P)$. In order to constrain the weights according to (\ref{OLS_constrain}), the available weights can be converted to time domain via IFFT, the last $M-L$ time steps are forced to zero, before the result is converted back to frequency domain by FFT. The input signal is then converted into the FFT domain as shown by (\ref{eq:fftInput}) and elementwise spectral multiplication (\ref{eq:spectralMult}) takes place in the FFT domain, before the output tensor is obtained with IFFT and the valid samples are saved and returned as the block output according to (\ref{eq:ifftOLS}).

\section{Multikernel Neural Network Model}
\label{sec:model}
Different application domains require different model structures, for instance, according to basic plant structures shown in Section \ref{sec:BSNLS}. Model representations of linear and nonlinear blocks therein were then provided by Sections \ref{sec:NFA} and \ref{sec:FIRblock}. We now form some basic model architectures with competence for multiplant representation and demonstrate their operation with respect to possibly different plant structures.

\subsection{Model Architectures}
\label{subsec:model_architecture}
\begin{figure}[!t]
  \centering
  \includegraphics[width=1\columnwidth]{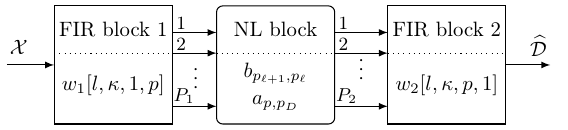}
  \caption{Block-structure of the FIR$_{P_1}$NL$_{P_2}$FIR model.}
  \label{fig:WHblockStructure}
\end{figure}

In order to introduce a compact notation of block-structured models, let us firstly consider the configuration shown in Fig.~\ref{fig:WHblockStructure} comprising of two FIR blocks and one NL block. This model architecture is here denoted by FIR$_{P_1}$NL$_{P_2}$FIR, indicating the order of applied FIR/NL blocks, while the subscript describes the number of connecting output channels. In our context, the number of inputs to the first block and the number of outputs of the last block is always one for single-input single-output (SISO) system identification problems.

At this point, we shall return to the fact that FIR blocks were introduced to primarily accomplish the desired multiplant representation. Now considering architectures with connected NL and FIR blocks, however, it should be noted that the multichannel interplay of NL and FIR blocks additionally implies the possibility of plant-specific nonlinear representation. This effect is due to the mathematical nature of nonlinear expansion, i.e., a NL block with multichannel output effectively spans a basis for nonlinear modelling, while actual nonlinear functions are achieved by the plant-specific aggregation with a subsequent FIR block.

Given the above short notation of block-structured model architectures, we can also refer to some important special cases of it. By discarding the first FIR block in Fig.~\ref{fig:WHblockStructure} and considering simply $P_1 = 1$ input channels, an NL$_{P_2}$FIR model architecture is obtained. With $P_2 = P > 1$ we then arrive at the \textit{parallel} Hammerstein model NL$_P$FIR, whereas $P_2 = 1$ yields the \textit{cascaded} Hammerstein model NL$_1$FIR, both of which renowned for instance in applications of acoustic system identification \cite{Stenger2000,kuech2006,Malik2012,Malik2013}. The cascaded model exhibits greatly less parameters, but appears more delicate in terms of the optimisation problem. Parallel models better learn the optimum parameters and achieve richer interplay of nonlinearity and memory by aggregating basis function with possibly different phase. By discarding the second FIR block in Fig.~\ref{fig:WHblockStructure}, we obtain an FIR$_{P_1}$NL Wiener architecture, where $P_1=P>1$ once more refers to a \textit{parallel} FIR$_P$NL and $P_1=1$ a \textit{cascaded} FIR$_1$NL representation.

\subsection{Model Parameter Identification}
\label{subsec:estimation}

Once the block structure of a model is defined based on available domain-knowledge, the optimal model parameter set $\vec{\Theta}^{[\kappa]} = \{  {{w}_1{[l,\kappa,1,p]}},{w}_2{[l,\kappa,p,1]}, {{a}_{p,p_D}}, {b}_{p_{\ell+1},p_{\ell}} \}$ needs to be determined for given a data set. To do so, the aim is to minimise the misalignment between the model output tensor ${\mathcal{D}} \big( \vec{\Theta}^{[\kappa]} \big)$ and the primary signal tensor $\mathcal{Y}$ as target, which can be stated as the optimisation problem
\begin{equation} \label{eq:costFct}
  \widehat{\vec{\Theta}}^{[\kappa]} = \argmin_{\vec{\Theta}} \Big\lVert \mathcal{Y} - {\mathcal{D}} \big( \vec{\Theta}^{[\kappa]} \big) \Big\rVert^2 \; ,
\end{equation}
i.e., the optimisation of the mean-square error (MSE) loss between model prediction and target. We here rely on the Adam optimiser \cite{KingmaB14} with learning rate $\mu_0 = 0.01$, $\beta_1 = 0.9$ and $\beta_2 = 0.999$. In every iteration all available data frames from all plants are employed for optimisation.

\subsection{Synthetic Multiplant Data}

Preparing for various applications, we employ computer-generated white Gaussian noise and recorded speech as input $x[n]$ for plant simulation. The speech signals are taken from the AEC Challenge data set \cite{Sridhar2021} each with a duration of $10$ seconds at a sampling frequency of $16\, \mathrm{kHz}$, while the Gaussian noise signals consist of $N=32,000$ samples per sequence.

Plants with Wiener and Hammerstein system structure according to Fig.~\ref{fig:sysId} are simulated. The plant-specific LTI behaviour therein is implemented by convolution
\begin{equation}
    \label{eq:LTI}
    d[n] = \sum_{l=0}^{511} h^{[\kappa]}[l] \, x[n-l] \;
\end{equation}
using plant-specific synthetic impulse responses $h^{[\kappa]}[l]$ drawn according to \cite{rir-generator, allen1979image}. For the nonlinear system part, two types of multiplant nonlinearity are employed:
\begin{itemize}
    \item a sigmoidal input-output characteristic 
\begin{equation} \label{eq:f_1}
    f^{[\kappa]}_\text{s}(x[n]) = \gamma^{[\kappa]} \, \arctan(\delta^{[\kappa]} \, x[n]) \; 
\end{equation}
with plant-specific scaling by $\gamma^{[\kappa]}$ and $\delta^{[\kappa]}$
\vspace*{1ex}
\item a hard-limiting input-output characteristic 
\begin{equation} \label{eq:f_2}
    f^{[\kappa]}_\text{c}(x[n]) = \begin{cases}
            x[n] & \text{for $\big\lvert x[n] \big\rvert \leq x_{\text{max}}[\kappa]$}\\
            x_{\text{max}}[\kappa]       &\text{otherwise } 
\end{cases} \;,
\end{equation}
where the clipping can be customised by $x_{\text{max}}[\kappa]$.
\end{itemize}
To quantify the respective nonlinearity, the 
\begin{equation}
    \mathrm{SDR}(x, f) = 10 \log_{10}\! \left[\frac{ E\{ ( \alpha \, x)^2 \} }{E\{ (f(x)- \alpha \, x)^2 \} } \right]
\end{equation}
of the linear part $\alpha\,x$ relative to the nonlinear residual of $f(x)$ is used, where $\alpha = { E\{ x^* \, f(x)  \}  }/{E\{  x^* \,x \}}$. 

\begin{figure}[!t]
  \centering
  \subfloat[nonlinear sigmoidal function.]{\includegraphics[width=0.88\columnwidth]{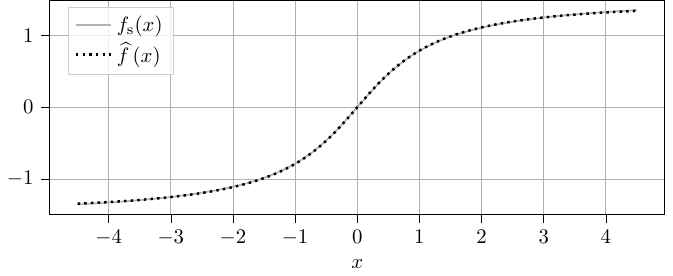}%
  \label{fig:hammerstein}}
  \hfil
  \subfloat[nonlinear clipping function.]{\includegraphics[width=0.88\columnwidth]{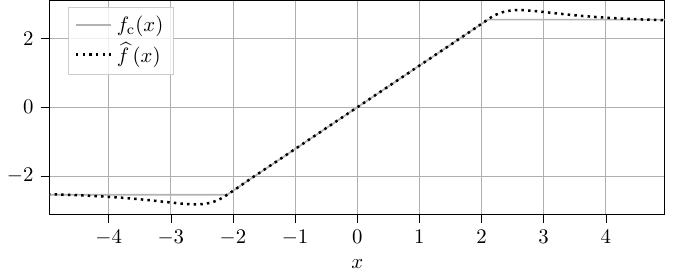}%
  \label{fig:wiener}}
  \caption{Memoryless plant nonlinearities and its identification by NL block.}
  \label{fig:NLs}
\end{figure}

A total of $N\!=\!10$ different plant realisations are generated for the Wiener and the Hammerstein scenario each, where the plant-specific nonlinearity ranges from SDR of $4 \,\mathrm{dB}$ to $32 \,\mathrm{dB}$ with mean SDR of $14\, \mathrm{dB}$. For the latter, as an example, the nonlinear functions are shown in Fig.~\ref{fig:NLs}. The verification in the following subsection takes place with $y[n] = d[n]$, i.e., the desired signal of $s[n]$ of Fig.~\ref{fig:SysId} being zero. From the simulated plant input and output sequences we segment frames according to (\ref{eq:frames}) with framesize $M= 2(L_1 + L_2)$ and frame-shift $R=(M-(L_1+L_2)+1)/2$, where $L_1$ and $L_2$ refer to the kernelsizes of the two FIR blocks, if they exist in the general model architecture of Fig~\ref{fig:WHblockStructure}. For the sake of completeness, the hyperparameters of the NL block therein are set to $D=5$ hidden layers with $P_\ell=6$ internal and $P=6$ output units.

\subsection{Verification of Multikernel Neural Network Models}

\begin{figure}[!t]
  \centering
  \subfloat[white noise excitation with $32,000 \cdot K$ time samples per epoch.]{\includegraphics[width=0.94\columnwidth]{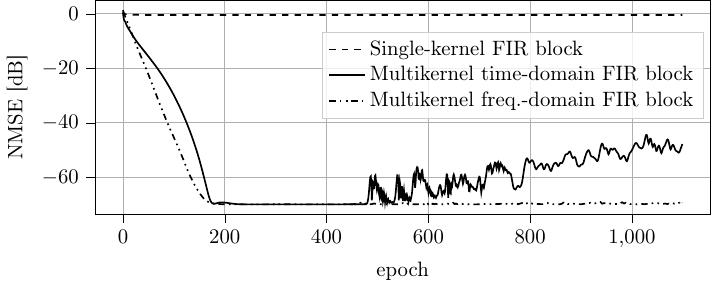}%
  \label{fig:wn_fir}}
  \hfil
  \subfloat[speech excitation with $160,000 \cdot K$ time samples per epoch.]{\includegraphics[width=0.94\columnwidth]{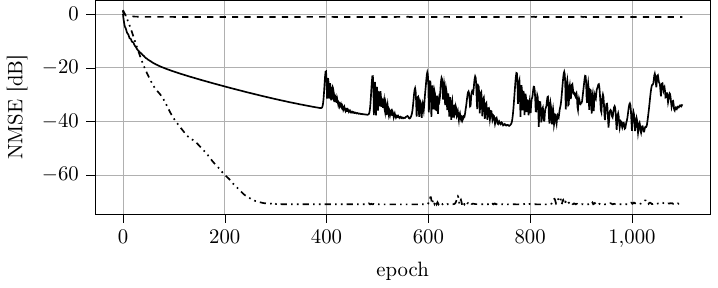}%
  \label{fig:speech_fir}}
  \caption{FIR-block modelling of linear multiplants with different input signals. 
  }
  \label{fig:FIR}
\end{figure}

Firstly, the FIR block and the NL block is considered for verification alone.
The FIR block in its variants of Section~\ref{sec:FIRblock} is employed for representation of the linear multiplant in \eqref{eq:LTI}. Fig.~\ref{fig:FIR} shows the learning curves of the FIR model in terms of normalised MSE for white Gaussian noise and speech-based data. In both cases a conventional single-kernel FIR block model clearly fails to match the model output to plant-specific observations provided as training data. The MSE remains high around $0$\,dB, demonstrating the need for multiple kernels in the model. For the white noise case in Fig. \ref{fig:wn_fir}, the multikernel FIR block then delivers MSE in the order of $-70 \,\mathrm{dB}$ which is successful and plausible for the synthetic data at hand. After $500$ training epochs, the NMSE of the time-domain FIR block scrapes off, but could be stabilised to $-70\,\mathrm{dB}$ with a smaller learning rate. With speech input into the plants, as shown by Fig.~\ref{fig:speech_fir}, the optimisation of the time-domain FIR block gets stuck around mediocre $-30\,\mathrm{dB}$ and merely the frequency-domain FIR block successfully attains the former $-70\,\mathrm{dB}$ NMSE. The self-correlation property of speech signals supposedly hinders efficient modelling in the time domain, but the frequency-domain ultimately rescues the training by analogy with adaptive filter theory \cite{Haykin2002}. The NL block alone of Section~\ref{sec:NFA} is then applied to nonlinear functions according to \eqref{eq:f_1} and \eqref{eq:f_2}. Fig.~\ref{fig:NLs} shows the successful nonlinear function approximation $\widehat{f}\,(x)$ of the respective nonlinearities, where the better approximation of the sigmoidal function is plausible with $\tanh$-activations inside the NL block.  

\begin{table*}[!t]
\renewcommand{\arraystretch}{1.2}
\setlength{\tabcolsep}{5.3pt}
\caption{Minimum NMSE $\mathrm{[dB]}$ achieved by using different model architecture on data from plants with different characteristics.}
\label{tab:verification}
\centering
\begin{tabular}{|c|c||c|c|c|c|c|c|c|c|c|}
\cline{3-11}
\multicolumn{2}{c||}{ \multirow{2}{*}{ }} & \multicolumn{9}{|c|}{\bf Model} \\
\cline{3-11}
\multicolumn{2}{c||}{} & \small FIR & \small NL$_1$FIR & \small NL$_6$FIR & \small FIR$_1$NL & \small FIR$_6$NL & \small FIR$_1$NL$_1$FIR &  \small FIR$_1$NL$_6$FIR & \small FIR$_6$NL$_1$FIR & \small FIR$_6$NL$_6$FIR  \\
\hline 
\multicolumn{2}{|c||}{\bf Wiener Data} & \multicolumn{5}{c|}{\multirow{2}{*}{ $L=512$ }}&\multicolumn{4}{c|}{\multirow{2}{*}{ $L_1=512\,$, $L_2=1$ }}  \\ 
\cline{1-2}
$\vec{h}$ & $f_\text{c}$& \multicolumn{5}{c|}{}  &\multicolumn{4}{c|}{}  \\  
\hline \hline
inv & inv& -11 & -11 & -11 & \bf -59 & \bf-52 & \bf-67 & \bf-60 & \bf-55 & \bf-57  \\
var & inv& -10 & -10 & -10 & \bf -59 & \bf-54 & \bf-61 & \bf-63 & \bf-44 & \bf-57  \\
inv & var& -13 & -13 & -13 & -16     & -16    & \bf-46 & \bf-59 & \bf-47 & \bf-44  \\
var & var& -12 & -12 & -12 & -14     & -14    & \bf-44 & \bf-60 & \bf-44 & \bf-42  \\
\hline
\hline
\multicolumn{2}{|c||}{\bf Hammerstein Data} & \multicolumn{5}{c|}{\multirow{2}{*}{ $L=512$ }}&\multicolumn{4}{c|}{\multirow{2}{*}{ $L_1=1\,$, $L_2=512$ }}  \\ 
\cline{1-2}
$f_\text{s}$&$\vec{h}$&  \multicolumn{5}{c|}{}  &\multicolumn{4}{c|}{}  \\
\hline \hline
inv&inv& -14 & \bf -65 & \bf -68 & -14 & -17 & \bf-68 & \bf-65 & \bf-67 & \bf-64  \\
var&inv& -7  & -22     & \bf -54 & -14 & -19 & \bf-65 & \bf-70 & \bf-64 & \bf-62  \\
inv&var& -14 & \bf -68 & \bf -67 & -7  & -10 & \bf-65 & \bf-63 & \bf-62 &\bf -60  \\
var&var& -7  & -23     & \bf -49 & -7  & -12 & \bf-49 & \bf-48 & \bf-49 & \bf-49  \\
\hline
\end{tabular}
\end{table*}

Next we conduct a comprehensive matrix study with the range of nonlinear model architectures described in Section~\ref{subsec:model_architecture} and a variation of data sets, i.e., based on single-plant (labelled ''inv'') and multi-plant (labelled ''var'') simulation of LTI and NL parts of Wiener and Hammerstein plant structures. The matrix is depicted in Table~\ref{tab:verification} including the final MSEs of the respective model parameter learnings. Just note that the FIR block here uses its time-domain version, since the study is restricted to the white Gaussian noise input. The simulation of Wiener data here uses clipping nonlinearity, while Hammerstein data is created sigmoidal.

A pure FIR model (left column) naturally fails the good MSE representation of the nonlinear data. For good nonlinear representation, however, it is still important where an additional NL block is placed in the architecture. The two-block NL$_P$FIR model, for instance, cannot represent the Wiener data and, conversely, the FIR$_P$NL model cannot represent Hammerstein data well (indicated by minimum NMSE in the order of $-12 \,\mathrm{dB}$). The FIR$_P$NL model is also not sufficient to match the Wiener data with multiplant nonlinearity. This requires an additional FIR block to follow the NL block as in the FIR$_P$NL$_P$FIR model (right column). Here, the plant-specific aggregation of basis functions into a plant-specific nonlinear representation takes place. Those models according to Fig.~\ref{fig:WHblockStructure} are sufficiently powerful to represent all constellations of the Wiener data such that the minimum NMSE attains $-40\, \mathrm{dB}$ and below. For Hammerstein data, already the two-block ''parallel'' NL$_P$FIR model with $P=6$ NL channels is very successful and can model all constellations with single- or multiplants. The larger three-block models (right) continue to be successful on the Hammerstein data as well.

\color{black}
We then consider established nonlinear system models, i.e.,
\begin{itemize}
    \item a ''memory polynomial'' \cite{morgan2006generalized} based on Eq.~\eqref{eq:powerApprox} but with FIR filters in place of the memoryless $a_p$ coefficients,
    \item a ''single-kernel'' neural Wiener-Hammerstein model~\cite{Halimeh2019}
\end{itemize}
for comparison with our strongest FIR$_6$NL$_6$FIR ''multikernel'' representation in Table \ref{tab:baseline}.
The memory polynomial uses a nonlinear model order $P=6$ and FIR filters of length $L=512$. It is practically linear in the parameters and solved in the least-squares (LS) sense per plant. The single-kernel neural baseline relies on the FIR$_6$NL$_6$FIR representation and is trained in the same framework with the multikernel version.
\color{black}
Apparently, the single-kernel model merely represents the data with a single plant in it (labelled ''inv'') well. The memory polynomial with its efficient individual LS solution per plant observation can naturally better represent multiplant nonlinearity (labelled ''var'') in Hammerstein data, but turns out to be limited to an average of $-27 \,\mathrm{dB}$ MSE, which can be traced to the cases of stronger nonlinearity in the data. The proposed multikernel neural network model can very good represent the Hammerstein data in all constellations as shown before. For the case of the Wiener data, the three-block multikernel model of this experiment is the only architecture to successfully match the data with low MSE. The memory polynomial with its structure of nonlinear basis functions followed by FIR filters does not fit the Wiener data.


\begin{table}[!t]
\renewcommand{\arraystretch}{1.2}
\setlength{\tabcolsep}{1.5pt}
\caption{Minimum NMSE $\mathrm{[dB]}$ of the proposed Multikernel neural network and baseline models.}
\label{tab:baseline}
\centering
\begin{tabular}{|c|c||c|c|c|}
\hline
\multicolumn{2}{|c||}{\bf Wiener Data} & \multicolumn{3}{c|}{\bf Model} \\ 
\hline
$\vec{h}$ & $f_\text{c}$ & \multirow{1}{*}{\small Multikernel} & \multirow{1}{*}{Memory Polynomial} & \multirow{1}{*}{Single-Kernel}  \\  
\hline \hline
inv & inv  & \bf-57 & -12 & \bf-48  \\
var & inv  & \bf-57 & -11 & -0  \\
inv & var  & \bf-44 & -10 & -8  \\
var & var  & \bf-42 & -10 & -1   \\
\hline
\hline
\multicolumn{2}{|c||}{\bf Hammerstein Data} &  \multicolumn{3}{c|}{\multirow{2}{*}{}} \\ 
\cline{1-2}
$f_\text{s}$ & $\vec{h} $  &  \multicolumn{3}{c|}{\multirow{2}{*}{}} \\
\hline \hline
 inv & inv & \bf-64 & \bf-55  & \bf-53 \\
 var & inv & \bf-62 &  -27  & -8 \\
 inv & var & \bf-60 &  \bf-55  & -0  \\
 var & var & \bf-49 &  -27  & -0  \\
\hline
\end{tabular}
\end{table}


\section{Application to Acoustic Echo Cancellation}
\label{sec:AEC}
The proposed framework is now applied in the domain of acoustic echo control. Acoustic echo appears as distraction in hands-free voice terminals. Due to acoustic coupling between loudspeaker and microphone, i.e., the echo path, the far-end receives a delayed version of the own voice. It inhibits fluent conversation during important double-talk speech periods.

\subsection{Hands-Free System}
\begin{figure}[!t]
  \centering
  \includegraphics[width=0.95\columnwidth]{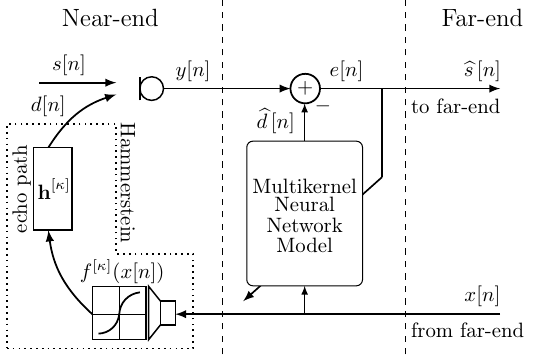}
  \caption{Setup for acoustic echo cancellation. The model output $\widehat{d}[n]$ approximates the echo signal $d[n]$, such that $e[n] \approx s[n]$ at the system output.}
  \label{fig:AEC}
\end{figure}
The near-end microphone of a hands-free speech system in Fig.~\ref{fig:AEC} captures the desired speech signal $s[n]$ at discrete time $n$ plus, unfortunately, a potentially nonlinear echo signal 
\begin{equation}
    d[n] = \sum_{m=1}^\infty h^{[\kappa]}[m] f^{[\kappa]}(x[n-m] )
\end{equation}
of the far-end (reference) speech $x[n]$. In this domain model, it is thus assumed that the echo path exhibits Hammerstein block structure. 
An acoustic echo canceller (i.e., generally in place of the multikernel network of Fig.~\ref{fig:AEC}) aims to eliminate the interfering echo $d[n]$ from the microphone signal $y[n] = s[n]+d[n]$ 
without distorting the desired signal $s[n]$.

More generally, systems for acoustic echo control traditionally consist of two stages. The acoustic echo canceller (AEC) is a first component and frequently models the echo path for echo estimation and subtraction by a linear FIR filter model. Echo path tracking in this system identification scenario is accomplished by adaptive filter algorithms, such as normalised least-mean squares (NLMS), recursive least-squares (RLS), or frequency-domain adaptive filters (FDAF) \cite{Haykin2002}. 
These algorithms are sensitive to local interference, such as double-talk, or echo path nonlinearities. The double-talk problem has been tackled successfully with adaptive learning rates based on double-talk detectors \cite{Benesty2001}, filter mismatch estimation \cite{valin2007}, or noisy state-space modelling and Kalman filtering \cite{Enzner06}. 
The problem of nonlinearities, which appear due to the loudspeaker in the echo path, has been addressed, for instance, by means of block-structured nonlinear Hammerstein models with fixed nonlinear basis functions as described in Section~\ref{subsec:LP}.
A different approach of dealing with nonlinearities is to apply additional spectral masking to the error signal $e[n]$ in Fig.~\ref{fig:AEC} in order to sufficiently suppress residual echo and potentially ambient noise.
This second stage is performed via statistical model-based postfilters \cite{VaryMartin2006,enzner2014acoustic} or neural networks \cite{halimeh_combining_2021,Valin2021,ivry_deep_2021,voit2023generalizedwiener}.
In this article, we are focusing on the isolated task of acoustic echo cancellation (AEC) based on the domain-specific Hammerstein echo path structure. It guides the corresponding model structure to include nonlinearities and in this way reduce residual echo below what a linear filter can do, without harm to the desired speech. The approach may still be extended with more complex block structures, additional postfilter or masking networks in future work.

\subsection{Domain-Specific Neural-Network for AEC}

For the AEC approach, certain requirements have to be met. While our frequency-domain FIR block is currently applicable for training conditions, the actual network application must further provide agility for double-talk robustness, non-stationary speech characteristic, as well as long and possibly time-varying room impulse responses (RIRs) easily in the order of $4000$ or more filter taps.
\begin{figure}[!t]
  \centering
  \subfloat[NL$_P$FIR for training.]{\includegraphics[width=0.45\columnwidth]{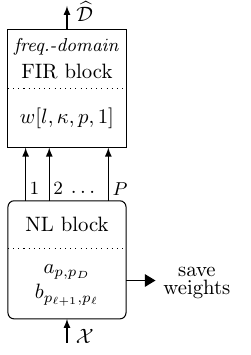}%
  \label{fig:training_model}}
  \hfil
  \subfloat[NL$_P$FDKF for testing.]{\includegraphics[width=0.45\columnwidth]{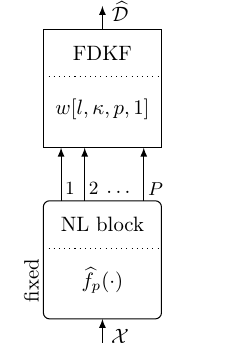}%
  \label{fig:testing_model}}
  \caption{Multikernel neural network models for training to extract nonlinear function across different plants, and testing used on unseen data with fixed nonlinear function approximation and FDKF as FIR block.}
  \label{fig:aec_models}
\end{figure}

To cope with the requirements we outline our training and testing strategy. We rely on the NL$_6$FIR model with frequency-domain FIR block as proposed in Section \ref{sec:model}. The NL block is specified with $D=3$ hidden layers with $P_\ell=9$ internal units and $P=6$ output units. While the training phase takes place in a far-end single-talk scenario, the test phase additionally involves double-talk scenes and uses completely unseen data, which was not included in the training. To manage, the testing hence employs the domain-specific frequency-domain adaptive Kalman filter (FDKF) \cite{Enzner06} and its multichannel version \cite{Malik2012} in place of the FIR block. The corresponding model architectures for training and test phases are shown in Fig.~\ref{fig:aec_models}. The training architecture in Fig.~\ref{fig:training_model} basically aims at the identification of a common nonlinear basis across plants, while the multikernel FIR block of the model here represents the typically variable acoustic impulse responses of a training data set. The weights of the NL block are saved when a minimal normalised MSE (NMSE) is reached. In the test phase, these weights are loaded into the corresponding NL block of the test architecture in Fig.~\ref{fig:testing_model}. With the FDKF in place of the FIR block, which is in the following termed the NL$_6$FDKF model, our intention is to cancel out nonlinear echo through double-talk with time-varying echo path impulse responses.

\subsection{Experimental Results}

Our experimentation uses the $16\,\mathrm{kHz}$ synthetic database provided by the ICASSP-2021 AEC Challenge \cite{Sridhar2021}. The structure of training and test data sets as well as far-end and near-end signals is retained in our experiments.

For first experiments we rely on own nonlinearities and echo path impulse responses. For training we generate $K=20$ 'self-made' echo path plants and feed with far-end signals $x[n]$ of $10$ seconds duration from the AEC Challenge training data set. The far-end signals are processed by nonlinear functions $f^{[\kappa]}(x)$ to mimic loudspeaker distortion, i.e., before the echo path. Specifically, the $\arctan$ defined in \eqref{eq:f_1} is used for plants $\kappa=1,\dots,5$, while the $\mathrm{clipping}$-nonlinearity given in \eqref{eq:f_2} is used for $\kappa=6,\dots,10$. By adjusting $\delta^{[\kappa]}$ or $x_\text{max}[\kappa]$, the nonlinearity is configured such that the $\mathrm{SDR}(x,f)$ ranges between $4\, \mathrm{dB}$ and $33 \,\mathrm{dB}$, whereby an average of $7 \,\mathrm{ dB}$ is achieved. Another ten linear samples ($f^{[\kappa]}(x)=x$, $\kappa = 11,\dots 20$) are included in the training set.
To create the echo signals $d[n]$, the nonlinearly mapped far-end signals are convolved with room impulse response $h^{[\kappa]}[n]$ of length $L=4096$ generated by the image method \cite{rir-generator, allen1979image} at different room positions. The near-end signal $s[n]$ is zero in all training samples.

\begin{figure}[!t]
  \centering
  \subfloat[single-talk scenario.]{\includegraphics[width=1\columnwidth]{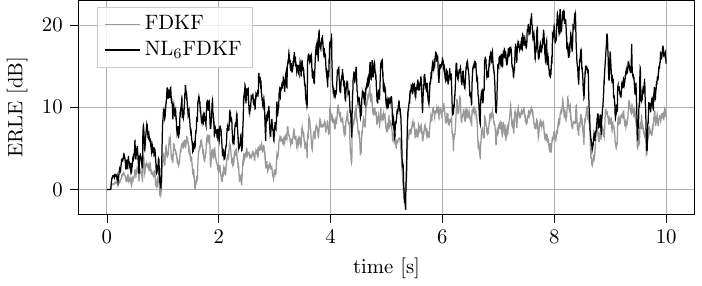}%
  \label{fig:erle_st}}
  \hfil
  \subfloat[double-talk scenario.]{\includegraphics[width=1\columnwidth]{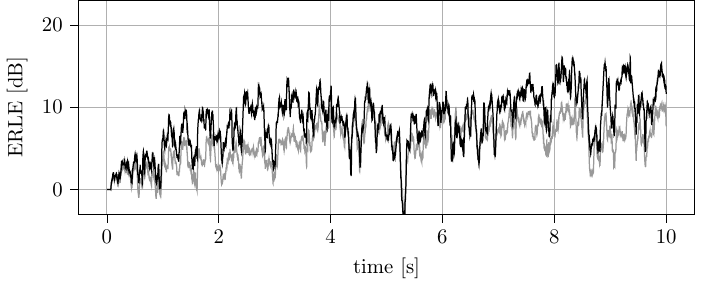}%
  \label{fig:erle_dt}}
  \hfil
  \subfloat[single-talk with echo-path-change scenario.]{\includegraphics[width=1\columnwidth]{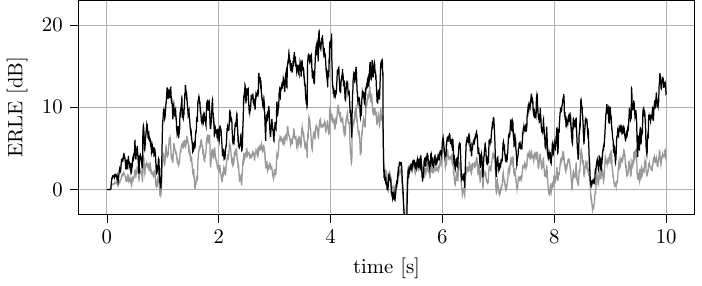}%
  \label{fig:erle_st_epc}}
  \caption{Test model from Fig. \ref{fig:testing_model} applied to 'self-made' test scenarios across $K=5$ echo plants with NL block trained on 'self-made' training data.}
  \label{fig:erle}
\end{figure}

For first testing, $K=5$ other echo plants are generated and fed with AEC Challenge far-end test samples. The procedure of creating the echo signals is similar to the training set, but here uses three other $\arctan$ and two other $\mathrm{clipping}$ nonlinearities with average $\mathrm{SDR}(x,f)= 7\,\mathrm{dB}$. The echo signals are further superimposed with the corresponding AEC Challenge near-end speech of $3-7$ seconds duration and zero-padding to $10$ seconds and scaling according to \cite{Sridhar2021}. Three different scenarios are considered: a) far-end single-talk similar to the training set; b) double-talk with near-end speech; c) far-end single-talk with echo-path change in the middle.

For AEC assessment, we then evaluate the $\mathrm{ERLE}[n] = 10\log_{10}\big(E\{ d^2[n] \} / E\{ ( d[n] - \widehat{d}[n])^2\}\big)$ averaged over the test data for the three different test scenarios. Fig.~\ref{fig:erle} depicts our NL$_6$FDKF results with respect to a baseline linear single-channel FDKF algorithm.
The superiority of the NL$_6$FDKF in comparison to the linear version is firstly visible in single-talk shown in Fig.~\ref{fig:erle_st}, where the linear FDKF is limited according the $\mathrm{SDR}(x,f)= 7\,\mathrm{dB}$, which NL$_6$FDKF can clearly overcome. The performance of NL$_6$FDKF is also seen in the double-talk scenario in Fig.~\ref{fig:erle_dt}, although the advantage here is smaller due to the near-end speech presence, mainly between second $3$ and $8$. If reconvergence is required after a change in the echo path, the NL$_6$FDKF is still above the linear FDKF, as can be seen in Fig.~\ref{fig:erle_st_epc}. 

\begin{figure}[!t]
  \centering
 \includegraphics[width=1\columnwidth]{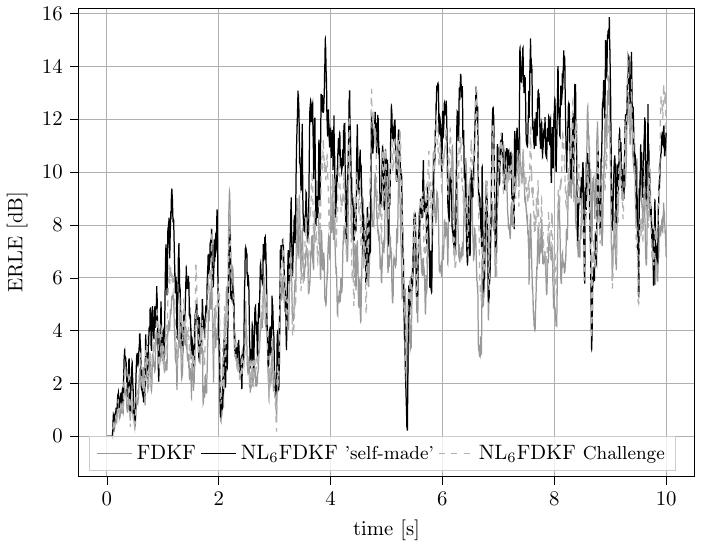}%
\vspace{-2ex}
\caption{AEC Challenge single-talk scenario. Averaged over $K=15$ randomly selected files from the AEC Challenge test set. NL block trained on 'self-made' training data and NL block trained on AEC Challenge data.}
  \label{fig:erle_st_icassp}
\end{figure} 

In order to investigate if the learned NL$_6$FIR nonlinearity also generalises to echo paths and nonlinearities originally used in the AEC Challenge, we further evaluate a subset of the original synthetic test set. It consists of $K=15$ randomly selected files from the test data set \cite{Sridhar2021}, comprising the respective far-end speech $x[n]$, the echo signal $d[n]$ including nonlinearity, and the near-end speech $s[n]$. Exact details of the echo signal generator are not known. Results of the previous linear FDKF and NL$_6$FDKF multikernel neural network model are shown in Fig.~\ref{fig:erle_st_icassp}. It can be seen that the 'self-made' NL$_6$FDKF partly addresses the nonlinearity of the original test data by some improvement over the just linear FDKF model. Additionally, an NL$_6$FDKF model is trained with original data from the AEC Challenge using a total of $K=40$ linear and nonlinear files. Surprisingly, this model trained on the supposedly more appropriate AEC Challenge data performs below the model trained on our 'self-made' data.


\section{Application to Wireless Self-Interference Cancellation}
\label{sec:SIC}
Full-duplex communication, where transmission and reception occur simultaneously on the same frequency resource, promises greater efficiency and flexibility in wireless systems. The primary challenge, however, is the self-interference (SI) of the transmitted signal leaking into the receiver and impeding the desired communication. In Wi-Fi, for instance, the transmitted signal power is around $20\,\mathrm{dB}$, while the information-bearing signal can be near the noise floor at about $-90\,\mathrm{dB}$. To avoid the obstructing interference, it must be cancelled by in the order of $110\,\mathrm{dB}$ from the receiver. Generally, the self-interference cancellation (SIC) hence involves techniques like passive or active shielding, analogue cancellation, and digital baseband adaptive cancellation to achieve the necessary reduction \cite{Heino_2015, Smida2023, Herd_2019}. In doing so, the strong self-interference is also subject to system nonlinearities in the SI path, which additionally challenges system modelling. In what follows, we briefly describe the wireless architecture, the related data for our study, the placement of a multikernel neural network for modelling, and the experimental results obtained with it.

\subsection{Wireless System}
\begin{figure}[!t]
  \centering
  \includegraphics[width=\columnwidth]{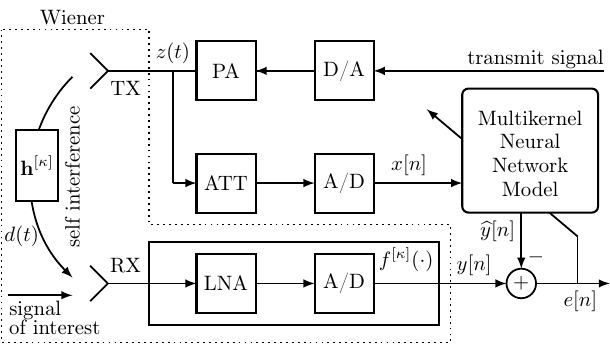}
  \caption{Wireless system with self-interference as a nonlinear Wiener model.}
  \label{fig:SIC}
\end{figure}

A configuration for cross-domain self-interference cancellation can be defined according to Fig.~\ref{fig:SIC}. The impact of power-amplifier (PA) nonlinearity is here effectively avoided by the utilisation of the analogue reference signal $z(t)$ in the continuous time domain $t$ to support the SIC according to previously proposed systems \cite{ahmed2015all,korpi2014reference}. This reference of the antenna output is attenuated (ATT) by auxiliary analogue components to the range of an A/D converter, before it can be supplied to a digital model of the SI path, here a multikernel neural network model. In this configuration, the SI path subject to modelling extends from the PA output $z(t)$ to the received signal $y[n]$ and thus refers to a Wiener plant model. The PA output firstly propagates through a  presumably linear TX to RX multiplant $h^{[\kappa]}(t)$ followed by nonlinear saturation in the LNA and A/D components due to its high power, i.e.,
\begin{equation}
y[n] = f^{[\kappa]}( h^{[\kappa]}(t) * z(t) ) \;,
\end{equation}
with the information-bearing signal being neglected in the scope of this paper. To achieve full-duplex operation, a wireless system must accurately model and cancel the overwhelming self-interference $y[n]$ from the transmitter into the receiver. Both the SI and the SIC are represented in the complex-valued baseband domain. Note that the reference $z(t)$ is in the SIC path represented by the digital version $x[n]$.

Experiments below will be performed for simulated high-throughput (HT) transmission in a wireless local area network (WLAN) with data available in \cite{Enzner_SICdata_Arxiv_2024}. It uses orthogonal frequency-division multiplexing (OFDM) baseband signals according to IEEE-802.11n with $20\, \mathrm{MHz}$ channel bandwidth. It will be relevant for our multikernel model that the nonlinearity of LNA and A/D is in the data realised with polar coordinates in the complex baseband domain as
\begin{equation}
\label{eq:fcomplex}
    f^{[\kappa]}(\cdot) = f_\text{mag}^{[\kappa]}(\lvert \cdot \rvert) \cdot \mathrm{e}^{\jmath \angle(\cdot)} \; 
\end{equation}
where $f_\text{mag}(\cdot)$ is a real-valued saturation and such that the average SDR of different plants $f^{[\kappa]}(\cdot)$ amounts to $10\,\mathrm{dB}$.

\subsection{Complex-valued Multikernel Neural Network}

\begin{figure*}[!t]
  \centering
  \includegraphics[width=0.94\textwidth]{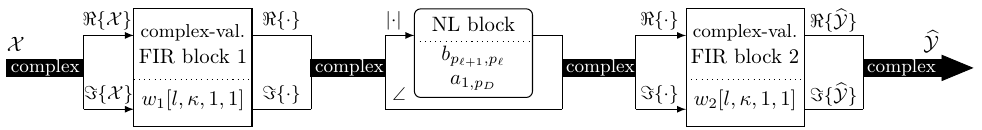}
  \caption{Block-structured FIR$_1$NL$_1$FIR model for complex-valued signals.}
  \label{fig:complexModels}
  \vspace{-1ex}
\end{figure*}

A multikernel FIR$_1$NL$_1$FIR model of Section~\ref{sec:model} is employed to represent the Wiener SI plant with its potential variability of linear and nonlinear components. Extension is required for applying this model in the complex-valued baseband domain. Fig.~\ref{fig:complexModels} shows the model architecture, where the bold links represent complex-valued signals, which are converted into Cartesian and polar forms as necessary. 

The complex-valued FIR blocks operate in Cartesian form, where the real and imaginary parts of the input signal, $\Re{\{\mathcal{X}\}}$ and $\Im{\{\mathcal{X}\}}$, are processed with two convolutional kernels such that valid components, $\Re\{\cdot\}$ and $\Im\{\cdot\}$, of a complex-valued output signal are delivered. In light of the plant information in \cite{Enzner_SICdata_Arxiv_2024}, our hyperparameters include the FIR filter lengths $L_1=20$, and $L_2=1$ samples in the time domain.

Taking into account the former expression \eqref{eq:fcomplex} for complex-valued nonlinearity in the data, we realise our complex-valued nonlinear modelling by means of the former NL block on the magnitude of a complex-valued signal and then recombine with the original phase, such that our model function $\widehat{f}(\cdot) = \widehat{f}_\mathrm{mag}(\lvert \cdot \rvert) \cdot \mathrm{e}^{\jmath \angle(\cdot)}$ complies with \eqref{eq:fcomplex}. The NL block is configured with $D = 3$ hidden layers with ${P}_\ell = 15$ internal nonlinear $\tanh$-activations each.

For the optimisation of the trainable parameters, the average mean-square error (MSE) of real and imaginary model output components w.r.t.\ the SI target $y[n]$ is used as the loss function. In the training phase all trainable parameters of FIR and NL blocks are adjusted for minimisation. In the test phase of the model, we retain the trained NL block parameters, while reoptimising the FIR block parameters in order to cope with new multiplants of the test data \cite{Enzner_SICdata_Arxiv_2024}.

\subsection{Experimental Results}

With the proposed multikernel model, a number of baseline models further take part in the evaluation, i.e.,
\begin{itemize}
    \item a multiplant 'linear FIR' model consisting of only one FIR block with kernelsize $L=20$,
    \item a plant-specific 'memory polynomial' model \cite{morgan2006generalized} with nonlinear order $P=15$ and filter length $L=20$,
    \item the proposed complex-valued multikernel FIR$_1$NL$_1$FIR neural network as shown by Fig.~\ref{fig:complexModels},
    \item and a complex-valued FIR$_1$NL$_1$FIR version with single kernel and otherwise the same configuration.
\end{itemize}

\begin{table}[!t]
\renewcommand{\arraystretch}{1.25}
\setlength{\tabcolsep}{3.3pt}
\caption{Minimum NMSE $\mathrm{[dB]}$ achieved with complex-valued models on complex-valued Wiener multiplant (labelled ''var'') data.}
\label{tab:sic}
\centering
\begin{tabular}{|c|c||c|c|c|c|}
\hline
\multicolumn{2}{|c|}{Model} & \multicolumn{1}{c|}{\small  FIR$_1$NL$_1$FIR }& \multicolumn{1}{c|}{\small FIR$_1$NL$_1$FIR} & {\small Memory} & {linear FIR}\\
\multicolumn{2}{|c|}{} & \multicolumn{1}{c|}{\small (Fig.\ref{fig:complexModels})}& \multicolumn{1}{c|}{\small (single kernel)} & {\small Polynomial} & {}\\
\hline\hline
\cline{1-2}
$\vec{h}$ & $f$ & & & & \\  
\hline 
inv & inv  & -69 & -44 & -14 & -11 \\
var & inv  & -70 & -5 & -13 & -12 \\
inv & var  & -68 & -25 & -14 & -12 \\
var & var  & -69 & -5 & -12 & -11  \\
\hline
\end{tabular}
\vspace{-1ex}
\end{table}

\begin{figure}[b] 
  \centering
  \includegraphics[width=\columnwidth]{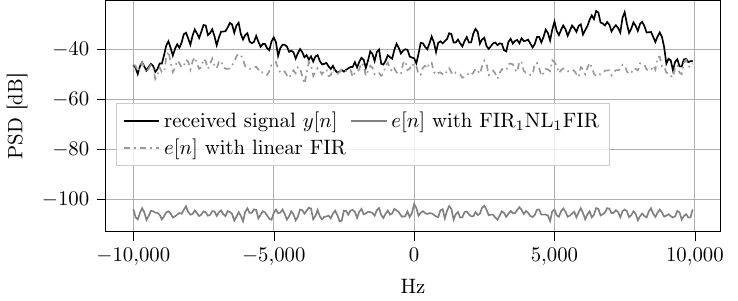}
  \caption{PSDs of received signal $y[n]$ without SIC and error signal $e[n] = y[n] - \widehat{y}[n]$ after SIC. Average of 10 signals from the ''inv/var'' test case.}
  \label{fig:sic_psds}
\end{figure}

Table \ref{tab:sic} shows the minimum NMSE achieved for the different models. The linear FIR model does not possess sufficient nonlinear modelling ability to represent the nonlinear Wiener data. The single-kernel FIR$_1$NL$_1$FIR model does not possess the capacity to represent multiplant data with either variable linear or nonlinear components (labelled ''var''). The memory polynomial rooted in Hammerstein modelling is neither able to represent the Wiener data with relevant accuracy. As a result, merely the multikernel model (left column) is able to model the SI data \cite{Enzner_SICdata_Arxiv_2024} well with a minimum MSE around $-70\,\mathrm{dB}$.

Fig. \ref{fig:sic_psds} finally provides an illustration of power spectral densities (PSDs) of the involved SI signals before and after cancellation. The unprocessed received signal $y[n]$ appears at the top and depicts an average attenuation level of $-40\,\mathrm{dB}$ according to passive SI shielding provided in the data. From there, a linear FIR model can just insignificantly reduce the SI to about $-50\,\mathrm{dB}$. The proposed multikernel FIR$_1$NL$_1$FIR model, however, creates an SIC residual below $-100\,\mathrm{dB}$, consisting of the incoming $-40\,\mathrm{dB}$ of the unprocessed data and the additional $-70\,\mathrm{dB}$ according to Table \ref{tab:sic}, and therefore attains the noise floor of the available data.


\section{Conclusion}
\label{sec:conclusion}

While deep learning is progressing relentlessly with powerful networks for classification and regression problems, it is not quite so commonly used for problems of nonlinear system identification. This paper has demonstrated conceptually and by considering typical applications of system identification that this might be rooted in the specific type of data and in the desire of custom-fit architectures. The data, if it comprises elements such as linear and nonlinear system responses, can be extremely inconsistent between different training samples and therefore complete hamper the identification of a good model representation. Regarding the model architectures, the system identification would typically rely on specific structures based on dedicated domain-knowledge. Yet, this paper demonstrates that the popular concept of architectures with trainable weights is applicable to nonlinear system identification as well, since there can be appropriate such model subsections with cross-data utility. Specifically, we have presented a block-structured approach for nonlinear modelling with a subset of trainable weights across the training data and another subset of plant-specific weights represented by multikernels. In this way the otherwise inconsistent data of different plant observations can be well represented both in training and testing, opening a deep learning perspective for nonlinear system identification.



\ifCLASSOPTIONcaptionsoff
  \newpage
\fi



\bibliographystyle{IEEEtran}
\bibliography{IEEEabrv,IEEEexample}

%









\end{document}